\documentclass{iopconfser}
\usepackage{amsmath}
\usepackage{amssymb}
\usepackage{amsfonts}
\usepackage{mathtools}
\usepackage{color}
\usepackage[mathscr]{euscript}

\newcommand{\dd}{{\rm d}}

\begin{document}

\title{Recent developments in semiclassical gravity}

\author{Benito A. Ju\'arez-Aubry}

\affil{Department of Mathematics, University of York, Heslington, York YO10 5DD, UK}

\email{benito.juarezaubry@york.ac.uk}

\begin{abstract}
Semiclassical gravity (SG) aims to describe the semiclassical regime of quantum gravity. In SG quantum fields curve classical spacetime in an effective way through the expectation value of their stress-energy tensor, while propagating in the spacetime they curve. I discuss some advances in SG in three directions: (i) in understanding structural properties or special solutions of SG in spacetimes with isometries or special Hadamard states (ii) in understanding the initial value formulation and (iii) in black hole evaporation and why information loss is not expected. Semiclassical frameworks beyond SG are briefly surveyed. The hope is that this short note will provide a good set of references and indicate interesting open problems and directions. 
\end{abstract}

\pagebreak

\section{Introduction}

In the absence of a fully workable and complete theory of quantum gravity, recent years have seen attention drawn to its semiclassical regime, described by semiclassical gravity (SG). While there is no clear-cut definition of the scale where this semiclassical regime lies, SG arguments are abundant in situations where both gravity and quantum phenomena are important. Notable examples are black hole radiation or the cosmology of the early universe. This is reasonable. As an analogy, semiclassical electrodynamics can be used to study the Lamb shift and spontaneous atomic decay without resorting to full quantum electrodynamics \cite{Crisp-Jaynes}. There are good reasons to think that something about quantum gravity can be learned from SG.

{\bf Notation:} We will denote quantum observables\footnote{A quantum observable here is an element of a non-commutative, unital $\star$-algebra of observables satisfying the axioms of an algebraic quantum field theory (QFT) in curved spacetimes, see e.g. \cite{Fewster-Verch-AQFT}.} and their distributional kernels with a boldface, e.g. ${\bf A}$, and the expectation value of an observable in a state, $\omega: {\bf A} \mapsto \omega({\bf A}) \in \mathbb{C}$, by $\langle {\bf A} \rangle_\omega$. Renormalised observables are denoted by $: {\bf A} :$ and their expectation values by $\langle :{\bf A}: \rangle_\omega$. 

The simplest matter model studied in SG is the free quantum Klein-Gordon theory. In the above notation, we denote the Klein-Gordon two-point function in the state $\omega$ by $\langle {\bf \Phi}(f) {\bf \Phi}(g) \rangle_\omega$, where $f, g \in C_0^\infty(M)$, and the expectation value of the stress-energy tensor in the state $\omega$ by $\langle :{\bf T}_{ab}:(f) \rangle_\omega$. A sufficient condition for $|\langle :{\bf T}_{ab}:(f) \rangle_\omega| < \infty$ is that $\omega$ be a Hadamard state, see \cite{Kay-Wald, Radzikowski} for details. In distributional kernel notation, we write $\langle {\bf \Phi}(x) {\bf \Phi}(x') \rangle_\omega$ (which is the Wightman function) and $\langle :{\bf T}_{ab}(x): \rangle_\omega$.

The semiclassical Einstein field equations with a Klein-Gordon field describe the gravitational field sourced by the expectation value of the stress-energy tensor of the Klein-Gordon field. Meanwhile, the quantum field obeys the Klein-Gordon equation in the spacetime it curves. The equations take the form
\begin{subequations}
\begin{align}
R_{ab}(x) - \frac{1}{2} R(x) g_{ab}(x) + \Lambda g_{ab}(x) = 8 \pi G_{\rm N} (\langle :{\bf T}_{ab}(x): \rangle_\omega + \alpha I_{ab}(x) + \beta J_{ab}(x)),  \label{EFE} \\ 
(-\Box + m^2 + \xi R(x))  \langle {\bf \Phi}(x) {\bf \Phi}(x') \rangle_\omega = (-\Box' + m^2 + \xi R(x')) \langle {\bf \Phi}(x) {\bf \Phi}(x') \rangle_\omega = 0, \label{KG}
\end{align}
\label{SEFE}
\end{subequations}
\noindent \!\!\! where $I_{ab} := R_{;ab} - \frac{1}{2} g_{ab} \Box R - \Box R_{ab} + \frac{1}{2} g_{ab} R^{cd} R_{cd} - 2 R^{cd} R_{cadb}$ and $J_{ab} := 2 R_{;ab} - 2 g_{ab} \Box R + \frac{1}{2}g_{ab} R^2 - 2 R R_{ab}$ are symmetric, covariantly conserved, rank-2 geometric tensors, and $\alpha$ and $\beta$ are dimensionless constant parameters, which account for renormalisation ambiguities in the definition of $:{\bf T}_{ab}:$. Throughout this note we focus on the semiclassical Einstein-Klein-Gordon system. Obvious generalisations include different matter fields, fermionic or bosonic, with or without gauge, linear or perturbatively interacting. However, System \eqref{SEFE} already captures the essence of the theory and makes transparent the difficulties in understanding its properties.

\noindent {\bf (I)} The first obvious complication is that Eq. \eqref{EFE} form a fourth order system in the spacetime metric. While this is not a major issue ---physicists and mathematicians have a large experience dealing with higher-order systems--- it does mean that the stability analyses of various important spacetimes currently understood in General Relativity (GR) do not carry over. One can expect 8 degrees of freedom in the gravitational sector of the theory. Upon using 4 diffeomorphisms, one should prescribe {\it a priori} 24 functions for the metric data ---the induced metric, second fundamental form and two additional $3\times3$ symmetric tensors. However, there are four constraints in the theory \cite{BAJAetal} and the contracted Bianchi identities, which can be used to fix 8 functions, leaving 16 free data (equals 8 degrees of freedom). 

\noindent {\bf (II)} The second complication that one faces is that System \eqref{SEFE} clearly has some non-local aspects. Indeed, the Klein-Gordon Wightman function of Eq. \eqref{KG} is a bi-distribution, $\langle {\bf \Phi}(\cdot) {\bf \Phi}(\cdot) \rangle_\omega \in \mathscr{D}'(M \times M)$, and thus outside of the scope of classical theorems of PDE theory. The existence of distinguished Green operators however alleviates some of our problems. Indeed, in a fixed globally hyperbolic spacetime, $(M, g)$, the initial value problem for the Wightman function is well-posed. Let $\langle \boldsymbol{\phi}(\cdot) \boldsymbol{\phi}(\cdot)  \rangle_{\omega_{\cal C}}$, $\langle \boldsymbol{\phi}(\cdot) \boldsymbol{\pi}(\cdot)  \rangle_{\omega_{\cal C}}$, $\langle \boldsymbol{\pi}(\cdot) \boldsymbol{\phi}(\cdot)  \rangle_{\omega_{\cal C}}$ and $\langle \boldsymbol{\pi}(\cdot) \boldsymbol{\pi}(\cdot)  \rangle_{\omega_{\cal C}}$ denote initial data for the Wightman function on a Cauchy surface, ${\cal C}$, in terms of correlations of CCR algebra fields $\boldsymbol{\phi}$ and $\boldsymbol{\pi}$ on ${\cal C}$. Let $E^{+/-}$ be the retarded/advanced Green operators of $-\Box + m^2 + \xi R$ and $E := E^- - E^+$ and let $u = \rho_0(E f):= (E f)|_{\mathcal{C}}$ and $v = \rho_1(E f):=(\nabla_n E f)|_{\mathcal{C}}$ be data for the solution $Ef$, $f \in C_0^\infty(M)$, of the classical Klein-Gordon equation. Then,
\begin{align}
\langle {\bf \Phi}(f) {\bf \Phi}(g) \rangle_\omega =  \langle \boldsymbol{\phi}(v) \boldsymbol{\phi}(v)  \rangle_{\omega_{\cal C}} - \langle \boldsymbol{\phi}(v) \boldsymbol{\pi}(u)  \rangle_{\omega_{\cal C}} - \langle \boldsymbol{\pi}(u) \boldsymbol{\phi}(v)  \rangle_{\omega_{\cal C}} + \langle \boldsymbol{\pi}(u) \boldsymbol{\pi}(u)  \rangle_{\omega_{\cal C}}
\label{KG-Sol}
\end{align}
solves Eq. \eqref{KG}, see e.g. \cite{Dimock} or \cite{JAMS} for details. 

\noindent {\bf (III)} The third issue that renders complicated System \eqref{SEFE} is connected to the previous one: renormalisation. Eq. \eqref{EFE} on its own looks like a local system of equations, but the source terms are the expectation value of a renormalised observable of the quantum theory, which is defined via the Hadamard subtraction procedure of a non-local expression. We illustrate this in the simpler case of $:{\bf \Phi}^2:$, which is defined as a locally covariant observable as $ :{\bf \Phi}^2:(f) \, := \int_{M\times M} \dd {\rm vol}_g \dd {\rm vol}_{g'} \left( {\bf \Phi}(x) {\bf \Phi}(x') - H_\ell(x,x') 1\!\!1 \right) f(x) \delta_{g'}(x,x'),$ 
where $f \in C_0^\infty(M)$ is supported in a convex normal neighbourhood in which $H_\ell$, which is the Hadamard parametrix for the Klein-Gordon equation at length scale $\ell$, admits the local form
\begin{align}
H_\ell(x,x') := \frac{1}{2 (2 \pi)^2} \left[ \frac{\Delta^{1/2}(x, x')}{\sigma_\epsilon(x, x')} +  v(x, x') \ln \left(\frac{\sigma_\epsilon(x, x')}{\ell^2} \right)  \right].
\end{align}

Here $\Delta$ is the van Vleck--Morette determinant and $v$ is a symmetric, smooth coefficient, formally determined by the so-called Hadamard recursion relations. The subindex $\epsilon$ in $\sigma_\epsilon$ denotes a suitable distributional regularisation for the Synge world-function.

While the source of the semiclassical Einstein equations \eqref{EFE} has local form, it comes from a non-local renormalisation procedure, which needs to be carried out as one {\it simultaneously} solves the field equations.

\noindent {\bf (IV)} The final complication is how to prescribe {\it Hadamard initial data} for the problem. In other words, how can one prescribe data for the two-point function, such that the solution consists of a globally hyperbolic spacetime and a Hadamard state $((M, g_{ab}), \omega)$? The main difficulty is that the Hadamard property cannot be easily formulated with {\it only} an initial achronal surface ---it is a {\it spacetime condition}.

We refer to these issues as ${\bf (I)}$-${\bf (IV)}$ in the text below.

Beyond all these mathematical issues, a question remains open on what is the class of Hadamard states for which semiclassical gravity is reliable. Coherent states are obvious candidates to be included. Ford and collaborators put forth a quantitative criterion for defining semiclassically reliable states \cite{Ford, Kuo-Ford, Ford-Wu}, which is indeed satisfied by coherent states. However, this criterion is unsatisfactory because it relies on a non-tensorial {\it correlation estimator}, often denoted by $\Delta_{abcd}(x,x')$ (the indices are not tensorial). The estimator seeks to capture the idea that quantum fluctuations should be small compared to the expectation value of the stress-energy tensor. However, this would arguably leave out important states, such as the Minkowski vacuum of free fields, which has vanishing expectation value. Recently, it has been shown that cat states, i.e. highly non-classical superpositions of coherent states, satisfy the aforementioned criterion \cite{Ahmedetal}. Whether this indicates that semiclassical gravity with cat states gives a reasonable description of physical phenomena, or a pathology of the criterion put forth in \cite{Kuo-Ford, Ford-Wu}, has not been clarified.

\section{Recent advances and symmetry reduced cases}

Semiclassical gravity solutions are known in maximally symmetric spacetimes, which greatly simplify the equations of motion \cite{BAJAcosmo, BAJA-ML, Gottschalketal}. Existence and uniqueness of solutions has been obtained in static spacetimes \cite{BAJAstat}. The case with spherical symmetry has been studied in detail in \cite{Sanders}. The relaxation to conformally static spacetimes with a conformal scalar has been studied in \cite{BAJA-Modak}.

Semiclassical cosmology has been studied with some intensity, and the problem has lent itself to some rigorous results, see \cite{Dappiaggietal, Eltzner-Gottschalk, Gottschalk-Siemssen, Gottschalketal2022, Meda-Pinamonti-Siemssen, Pinamonti, Pinamonti-Siemssen} for recent progress. It should be mentioned that \cite{BAJAcosmo, Gottschalketal} in de Sitter spacetime also have cosmological motivations and, in particular, \cite{BAJAcosmo} is motivated by the cosmological constant problem. In terms of methods, a technique is introduced in \cite{Gottschalk-Siemssen}, which writes the semiclassical cosmological equations as an infinite tower of so-called {\it moment equations}. Existence of FLRW solutions is shown by this method. An interesting question is whether this framework can be generalised beyond cosmology. It is noteworthy that local existence and uniqueness in semiclassical cosmology has been obtained in \cite{Meda-Pinamonti-Siemssen} within the FLRW class.

Local existence and uniqueness in spacetimes without special isometries has been recently obtained within a very special class of two-point functions, which have the feature that they yield a renormalised stress-energy tensor expectation value proportional to the spacetime metric, greatly simplifying the semiclassical Einstein equations \cite{BAJA2024}.

The problem of back-reaction has also been studied, e.g. in maximally symmetric spacetimes, lower dimensions (where SG is of second order) or black holes, see e.g. \cite{Thompson-Winstanley, Casalsetal, Medaetal, delRio}.

A singularity theorem with weakened hypotheses, appropriate to SG, has been put forth by Fewster and Kontou in \cite{Fewster-Kontou}. A generalisation of the Hawking black hole area theorem in SG by Kontou and Sacchi appears in \cite{Kontou-Sacchi}.

\section{The initial value problem and structural properties}

The issues raised in ${\bf (I)}$ have been a source of concern since the 90s. Notably, Parker and Simon \cite{Parker-Simon} applied Simon's {\it reduction of order} to discard potential runaway solutions (growing exponentially fast in time). It has been argued however that this could leave out interesting solutions of the theory \cite{Flanagan-Wald}.

A recent paper by Ju\'arez-Aubry, Kay, Miramontes and Sudarsky \cite{BAJAetal} conjectures that physical solutions to the theory have smooth dependence in $\hbar$ (equivalently $\ell_{\rm Planck}^2$) at $\hbar = 0$. Amongst these solutions, the ones with analytic dependence can be written as convergent series in $\hbar$.\footnote{An undesirable feature here is that $\hbar$ is dimensionful. However, fixing a renormalisation scale, $\ell$, the combination $\hbar/\ell^2$ is a suitable dimensionless perturbation parameter.} It is further conjectured that for such physical solutions the initial value problem reduces to second order ---this intuition is formed by thinking of the $\hbar$-analytic case, and inspired by \cite{Parker-Simon}.

Furthermore, \cite{BAJAetal} offers conjectures on the well-posedness of the full semiclassical gravity theory, together with a definition of {\it Hadamard initial data}, which encompasses {\it a priori} an infinite tower of initial data constraints obtained from the semiclassical Einstein equation on the initial surface and derivatives thereof. A finite cut-off of this constraints tower provides initial data for adiabatic states. These constraints are however not only very challenging to solve, but even {\it computing the expressions to be solved} is very difficult ---though possible--- because of {\bf (III)} and {\bf (IV)}, as exemplified in \cite{BAJAetal2}.

The characteristic initial value problem might simplify analysis of the initial data of SG because problem {\bf (IV)} might be more manageable. On this note, Janssen and Verch explored the problem of Hadamard renormalisation on a light-cone in spacetimes with spherical symmetry in \cite{Janssen-Verch}. The construction of Hadamard states from characteristic data has received attention since the 80s, starting with Kay's {\it magic} formula for Hadamard states on a null surface \cite{Dimock-Kay, Kay-Wald}. Recent studies include e.g. \cite{DMP, Gerard-Wrochna}.

Concerning the stability of SG, Meda and Pinamonti have studied linear stability in a semiclassical scalars toy model, providing criteria for avoiding runaway solutions \cite{Meda-Pinamonti}. This model generalises a perturbatively solvable semiclassical scalars model of Ju\'arez-Aubry, Miramontes and Sudarsky \cite{JAMS}.

Challenge {\bf (II)} remains hard to penetrate, but some toy models might help make progress on this front. It is worth mentioning here the very recent pre-print \cite{Finster-Murro-Schmid}.

\section{The black hole information loss puzzle}

Semi-classical arguments led Hawking to suggest that black holes evaporate \cite{Hawking}. However, he argued, in the evaporation process information is lost. The situation is summarised in the famous black hole evaporation conformal diagram of Fig. 5 in \cite{Hawking}. This conclusion suggests that there are physical processes in nature that violate quantum and classical predictability; the former is better known as unitarity.

In a recent paper \cite{BAJAinfoloss}, Ju\'arez-Aubry has argued that this conclusion neglects the abundant evidence that QFT generically renders Cauchy horizons unstable under semiclassical back-reaction (see \cite[Sec. 2.2]{BAJAinfoloss} and the notable examples \cite{Hollandsetal1, Hollandsetal2}), pointing out that the black hole evaporation `endpoint' breaks global hyperbolicity. In \cite{BAJAinfoloss}, `Quantum strong cosmic censorship' conjectures are put forth, summarised as follows:

\begin{enumerate}
\item For a QFT in a fixed spacetime with a Cauchy horizon, ${\cal C}$, with respect to the domain of dependence, ${\cal D}({\cal S})$, of a (non-Cauchy) achronal surface, ${\cal S}$, it is generically impossible to extend a Hadamard state on ${\cal D}({\cal S})$ to and beyond the Cauchy horizon as a Hadamard state.
\item Cauchy horizons are generically unstable in SG (due to a breakdown of the Hadamard property).
\end{enumerate}

If these conjectures hold, SG actually predicts a final singularity, which emanates from the evaporation endpoint, where spacetime terminates. However, one expects that quantum gravity should cure this pathology. In this strong-cosmic-censored picture, there is no post-evaporation region, hence no loss of information. See again \cite{BAJAinfoloss}, especially Sec. 3 and 4, for details.

\section{Other semiclassical gravitational frameworks}

This section briefly comments on attempts to go beyond SG, which the reader might find interesting. A common theme is the incorporation of stochastic dynamics. 

Stochastic gravity \cite{Hu-Verdaguer} seeks to incorporate quantum fluctuations of the matter fields into the standard semiclassical framework. The gravitational equations take the form of an Einstein-Langevin equation.

Oppenheim recently introduced a stochastic theory of classical spacetime coupled to quantum matter \cite{Oppenheim}, dubbed {\it post-quantum}, which is {\it a priori} agnostic about whether gravity is fundamentally quantum.

Mathematical aspects of the question of integrating dynamical quantum collapse models into SG have been explored in the literature, see \cite{BAJAKS, BAJAetal}. This programme has the risk of prompting superluminal signalling \cite{Chungetal}. Recent schematic ideas attempting to circumvent this issue are proposed in \cite{Mucinoetal}.\footnote{It has been argued, see e.g. \cite{Grossardt}, that state collapse can render SG physically consistent at fundamental level , but the mathematical consistency of this framework remains an open question \cite{BAJAetal}.} In these approaches, semiclassical gravity becomes `suspended' during state collapse. Another proposal relying on the `suspension' of SG has been very recently put forth by Pipa. Here instead the SG equations are switched on or off depending on certain networks of interactions between matter fields \cite{Pipa}.


\section*{Acknowledgments}

This work is supported by the EPSRC Open Fellowship EP/Y014510/1. This paper is based on the talk `Advances in Semiclassical Gravity', presented in Session D4 of the GR24 Conference at Glasgow in the Summer of 2025. Warm thanks are due to Chris Fewster for some helpful comments.

\end{document}